\documentclass[12pt]{article}
\usepackage{amssymb,amsmath}
  \newcommand{\Fc}{\mathcal{F}}

\newcommand{\pd}{\partial}
\newcommand{\ba}{\begin{eqnarray}}
\newcommand{\ea}{\end{eqnarray}}
\newcommand{\be}{\begin{equation}}
\newcommand{\ee}{\end{equation}}

\begin{document}
\title{\textbf{On Bouncing Solutions in Non-local Gravity}}

\author{Alexey~S.~Koshelev$^{1}$\footnote{Postdoctoral researcher of
FWO-Vlaanderen, E-mail:
alexey.koshelev@vub.ac.be} \ and \ Sergey~Yu.~Vernov$^{2,3}$\footnote{E-mail: vernov@ieec.uab.es, svernov@theory.sinp.msu.ru} \vspace*{3mm} \\
{\small $^1$Theoretische Natuurkunde, Vrije Universiteit Brussel and The
International Solvay Institutes,}\\{\small  Pleinlaan 2, B-1050 Brussels,
Belgium}\\
{\small $^2$Instituto de Ciencias del Espacio (ICE/CSIC) and} \\
{\small  Institut d'Estudis Espacials de Catalunya (IEEC),}\\
{\small Camp. UAB, Fac. Ci\`encies, T. C5,
 E-08193, Bellaterra, Barcelona, Spain}\\
{\small $^3$Skobeltsyn Institute of Nuclear Physics, Lomonosov Moscow State University,}\\
{\small Leninskie Gory 1, 119991, Moscow, Russia}}
\date{ }
\maketitle

\vspace{-7.2mm}
\begin{abstract}
A non-local modified gravity model with an analytic function of the d'Alembert
operator is considered.
This model has been recently proposed as a possible way of resolving the
singularities problem in
cosmology. We present an exact bouncing solution, which is simpler compared to
the already known one in this model in the sense it does not
require an additional matter to satisfy all the gravitational equations.
\end{abstract}

\section{Introduction}

Modified gravity cosmological models have been proposed with the hope to
find resolutions to the important open problems of the standard cosmological
model. One possible modification
which allows to improve the ultraviolet behavior and even to get
a renormalizable theory of quantum gravity is
the addition of higher-derivative terms to the Einstein--Hilbert action (as one
of the
first papers we can mention~\cite{Stelle}).
Unfortunately, models with the higher-derivative terms have ghosts. A possible
way to
overcome this problem is to consider
a non-local gravity.

The main theoretical motivation for
studying cosmological models with the non-local
corrections to the Einstein-Hilbert action comes from the string field theory
\cite{sft_review}. These corrections
usually contain exponential functions of the d'Alembertian operator and
appear in such stringy models as effective tachyonic actions. The
majority of the non-local cosmological models motivated by such
structures explicitly include an analytic or meromorphic
function of the d'Alembert operator~\cite{NLG,NLScF,BMS,BKM,AJV0711,ExactNLG,BGKM}.

Usually both the general relativity and the modified gravity models are
described by
a nonintegrable system of equations and only particular exact solutions can be
obtained. At the same time exact solutions play an important role in
the cosmological models since one must consider perturbations in order to
claim the model is realistic. Needless to say exact solutions for
non-local nonlinear equations is an extremely tough subject. Some
studies for nonlocal gravitational models with exact solutions can be found
in~\cite{BMS,BKM,AJV0711,ExactNLG}.

\section{Action and Equations of Motion}
The nonlocal modification of the Einstein gravity, which has been
proposed in~\cite{BMS,BKM}, is described by the following action:
\begin{equation}
 S=\int d^4x\sqrt{-g}\left(\frac
 {M_P^2}{2}R+\frac{1}{2}R\Fc(\Box/M_*^2)R-\Lambda\right),
 \label{nlg_action}
\end{equation}
where   $M_P$ is the Planck mass.
 $M_{\ast}$ is the mass scale at which the higher derivative
terms in the action become important.
An analytic function
$\Fc(\Box/M_*^2)=\sum\limits_{n\geqslant0}f_n\Box^n$ is an ingredient
inspired by the SFT. The operator $\Box$ is the covariant
d'Alembertian. In the case of an infinite series we have a non-local
action.

Let us introduce dimensionless coordinates
$\bar{x}_\mu=M_* x_\mu$ and $\bar{M}_P=M_P/M_*$. It is easy to see that $\Fc(\Box/M_*^2)=\Fc(\bar{\Box})$, where $\bar{\Box}$ is
the  d'Alembertian in terms of dimensionless coordinates. In the following we
omit bars using only dimensionless
coordinates.

A straightforward variation of action (\ref{nlg_action}) yields the
following equations
\begin{equation}
\begin{split}
&(M_P^2+2\Fc(\Box)R)\left(R_{\mu\nu}-\frac{1}{2}Rg_{\mu\nu}\right)=2(D_\mu\pd_\nu-g_{\mu\nu}
\Box)\Fc(\Box) R-\Lambda g_{\mu\nu}+{}\\
&{}+\frac{1}{2}\sum_{n=1} ^\infty f_n\sum_{l=0}^{n-1}\Bigl[\pd_\mu\Box^l
R  \pd_\nu\Box^{n-l-1}  R
+\pd_\nu\Box^l  R  \pd_\mu\Box^{n-l-1}  R  -{}\\
&{}-g_{\mu\nu}\left(g^{\rho\sigma} \pd_\rho\Box^l  R
\pd_\sigma\Box^{n-l-1}  R  +\Box^l  R  \Box^{n-l}  R
\right)\Bigr]-\frac{1}{2}
 R \Fc(\Box) Rg_{\mu\nu}\,,
\end{split}
\label{eqEinsteinRonly}
\end{equation}
where $D_\mu$ is the covariant derivative. It is useful~\cite{BKM} to
write down
the trace equation:
\begin{equation}
\begin{split}
M_P^2R-&\sum_{n=1}^\infty f_n\sum_{l=0}^{n-1}\Bigl(\pd_\mu\Box^l  R
\pd^\mu\Box^{n-l-1}  R +2\Box^l  R  \Box^{n-l}  R\Bigr)-
6\Box\Fc(\Box)R=4\Lambda\,.
\end{split}
\label{eqEinsteinRonlytrace}
\end{equation}
\section{General Ansatz for finding Exact Solutions}

It has been shown in~\cite{BMS} that the following ansatz
\begin{equation}
\Box R=r_1R+r_2\,,
 \label{ansatz2}
\end{equation}
with $r_1\neq 0$, is useful in finding exact solutions. Using
(\ref{ansatz2}), the trace equation becomes
\begin{equation}
A_1R+A_2\left(2r_1R^2+ \partial_\mu R \partial^\mu R\right)+A_3=0\,,
\label{Trequ_anzats}
\end{equation}
where
\begin{equation*}
\begin{split}
A_1&=-M_P^2+4\Fc'(r_1)r_2-2\frac{r_2}{r_1}(\Fc(r_1)-f_0)+6\Fc(r_1)r_1\,,\qquad
A_2=\Fc'(r_1),\\
A_3&=4\Lambda+\frac{r_2}{r_1}M_P^2+\frac{r_2}{r_1}A_1-2\frac{r_2^2}{r_1}\Fc'(r_1).
\end{split}
\end{equation*}

The simplest way to get a solution to equation (\ref{Trequ_anzats}) is
to put all the above coefficients to zero. Relations $A_j=0$, $j=1,2,3$ determine values of $r_1$, $r_2$ and
also fix the
cosmological constant:
\begin{equation}
\label{r12} \Fc'(r_1)=0,\qquad
r_2={}-\frac{r_1[M_P^2-6\Fc(r_1)r_1]}{2[\Fc(r_1)-f_0]},\qquad
\Lambda=-\frac{r_2M_P^2}{4r_1}=M_P^2\frac{[M_P^2-6\Fc(r_1)r_1]}{
8[\Fc(r_1)-f_0]}.
\end{equation}

\section{Exact solutions and their applications}

Let us consider solutions in the spatially flat
Friedmann--Lema\^{i}tre--Robertson--Walker (FLRW)
metric with the interval
$ds^2={}-dt^2+a^2(t)\left(dx_1^2+dx_2^2+dx_3^2\right)$.

The very important result for this kind of models was a construction of
an analytic solution describing the non-singular bounce
\begin{equation}
 a(t)=a_0\cosh(\lambda t),
 \label{exact_scalefactor_a}
\end{equation}
where $a_0$ is an arbitrary constant and $\lambda=\sqrt{\Lambda}/3M_P$.
To satisfy all equations (\ref{eqEinsteinRonly}) one should add some radiation
to the model. This is the exact
analytic result and we refer the readers to \cite{BMS,BKM} about all the
details.

Let us consider another solution, which satisfies the ansatz
(\ref{ansatz2}). Namely,
\begin{equation}
a(t)=a_0e^{\frac{\lambda}2 t^2},
 \label{sol2}
\end{equation}
where $a_0$ is an arbitrary constant. On this solution
\begin{equation}
H(t)=\lambda t,\quad R=12\lambda^2 t^2+6\lambda,\quad
\Box R=-72\lambda^3 t^2-24\lambda^2, \quad \Rightarrow\quad
r_1=-6\lambda,\quad r_2=12\lambda^2,
 \label{sol2HRboxR}
\end{equation}
where $H=\dot{a}/a$ is the Hubble
parameter and the denotes the differentiation with respect to time $t$.
From the condition $A_3=0$ we get
$\Lambda=\lambda M_P^2/2$.
From $A_1=0$ and $A_2=0$ we get the following constraints on the function $\Fc$ and its first derivative
in the point $r_1$:
\begin{equation}
\Fc(r_1)={}-\frac{M_P^2}{32\lambda}-\frac{f_0}{8}\,,  \qquad \Fc'(r_1)=0\,.
\end{equation}

There are two independent Einstein equations in the FLRW metric. Let us consider
''00'' component of system (\ref{eqEinsteinRonly}), which, after imposing
the
simplifying ansatz and using conditions $A_j=0$, reads
as
the second order differential equation for the Hubble parameter
$H(t)$:
\begin{equation}
\Fc(r_1)\left[H\ddot H +3H^2\dot H-\frac{1}{2}{\dot
H}^2+\frac{r_1}{2}H^2+\frac{r_2}{24}\right] =0\,.
\label{FRW_eqEinsteinRonlyansatztt}
\end{equation}
Substituting (\ref{sol2HRboxR}), we obtain that equation
(\ref{FRW_eqEinsteinRonlyansatztt}) is satisfied,
so function (\ref{sol2}) is a solution to all Einstein equations. Note that
we do not add any matter to get the exact solution.

We stress that a construction of exact solutions is obviously a non-trivial
task
and to the moment only one non-trivial exact analytic bouncing solution
(\ref{exact_scalefactor_a}) in the class of models given by action
(\ref{nlg_action}) is analyzed~\cite{BKM}. We present here another bouncing
solution (\ref{sol2}) which
is simpler compared to (\ref{exact_scalefactor_a}) in the sense it does not
require an additional matter to be present.

We are leaving open the questions of perturbation spectrum for exact solutions
in this nonlocal model and those
applications to describing the bounce phase and the initial inflation stage.
These question will be addressed in the forthcoming publications.

$~$

\textbf{Acknowledgments.} The authors
are grateful to the organizers of the Dubna International
SQS'11 Workshop for the hospitality and the financial
support. The authors thank T. Biswas for very useful and
stimulating discussions.
The work is supported
in part by the RFBR grant 11-01-00894. A.K. is supported in part by
the Belgian Federal Science Policy Office through the Interuniversity
Attraction
Pole P6/11, and in part by the ``FWO-Vlaanderen'' through the project
G.0114.10N.
 S.V. is supported in part by
grants of the Russian Ministry of Education and Science NSh-4142.2010.2 and NSh-3920.2012.2
and by contract CPAN10-PD12 (ICE, Barcelona, Spain).

\end{document}